\documentclass[conference]{IEEEtran}
\IEEEoverridecommandlockouts
\usepackage{cite}
\usepackage{amsmath,amssymb,amsfonts}
\usepackage{algorithmic}
\usepackage{graphicx}
\usepackage{textcomp}

\usepackage{xcolor}
\definecolor{Black}{RGB}{0, 0, 0}
\definecolor{Orange}{RGB}{230, 159, 0}
\definecolor{SkyBlue}{RGB}{86, 180, 233}
\definecolor{BluishGreen}{RGB}{0, 158, 115}
\definecolor{Yellow}{RGB}{240, 228, 66}
\definecolor{Blue}{RGB}{0, 114, 178}
\definecolor{Vermillion}{RGB}{213, 94, 0}
\definecolor{ReddishPurple}{RGB}{204, 121, 167}
\definecolor{ukpurple}{RGB}{199, 16, 92}

\usepackage[inline, shortlabels]{enumitem}

\usepackage{units}

\usepackage{amsthm}
\theoremstyle{definition}
\newtheorem{problem}{Problem}

\usepackage{pgfplots}
\pgfplotsset{compat=1.18}
\usepgfplotslibrary{fillbetween}

\usepackage{tabularx}
\usepackage{booktabs}
\usepackage{hyperref}
\usepackage{orcidlink}
\usepackage[capitalise]{cleveref}
\crefname{problem}{Problem}{Problems}

\usepackage{acronym}
\acrodef{mpc}[MPC]{model predictive control}
\acrodef{ai}[AI]{artificial intelligence}
\acrodef{ml}[ML]{machine learning}
\acrodef{opf}[OPF]{optimal power flow}
\acrodef{hilp}[HILP]{high-impact, low-probability}
\acrodef{arima}[ARIMA]{autoregressive integrated moving average}
\acrodef{arima}[ARIMA]{autoregressive integrated moving average}
\acrodef{knn}[$k$-NN]{$k$-nearest neighbors}
\acrodef{svm}[SVM]{support vector machine}
\acrodef{rnn}[RNN]{recurrent neural network}
\acrodef{io}[I/O]{input/output}
\acrodef{facts}[FACTS]{flexible AC transmission system}
\acrodef{lpv}[LPV]{linear-parameter varying}
\acrodef{cnn}[CNN]{convolutional neural network}

\usepackage{enumitem}
\newenvironment{compactitem}{
	\vspace{-0.65em}
	\begin{itemize}[noitemsep, topsep=0pt, parsep=0pt, partopsep=0em, leftmargin=1.1em]
	}{
	\end{itemize}
	\phantom{a}
	\vspace{-1.1em}
	}

\usepackage{todonotes}
\def\BibTeX{{\rm B\kern-.05em{\sc i\kern-.025em b}\kern-.08em
    T\kern-.1667em\lower.7ex\hbox{E}\kern-.125emX}}
\begin{document}

\title{
Predictions and Decision Making for \\ Resilient Intelligent Sustainable Energy Systems
}

\author{
\IEEEauthorblockN{%
Martin Braun\IEEEauthorrefmark{4}\IEEEauthorrefmark{2},
Christian Gruhl\IEEEauthorrefmark{4},
Christian A. Hans\IEEEauthorrefmark{4}\IEEEauthorrefmark{1},
Philipp Härtel\IEEEauthorrefmark{4}\IEEEauthorrefmark{2},
Christoph Scholz\IEEEauthorrefmark{4}\IEEEauthorrefmark{2},\\
Bernhard Sick\IEEEauthorrefmark{4}, 
Malte Siefert\IEEEauthorrefmark{2},
Florian Steinke\IEEEauthorrefmark{3},
Olaf Stursberg\IEEEauthorrefmark{4},
Sebastian Wende-von Berg\IEEEauthorrefmark{4}\IEEEauthorrefmark{2}
}
\IEEEauthorblockA{\IEEEauthorrefmark{4}University of Kassel, Germany. \emph{\{%
\href{mailto:martin.braun@uni-kassel.de}{martin.braun},
\href{mailto:cgruhl@uni-kassel.de}{cgruhl},
\href{mailto:hans@uni-kassel.de}{hans},
\href{mailto:bsick@uni-kassel.de}{bsick},
\href{mailto:stursberg@uni-kassel.de}{stursberg}%
\}@uni-kassel.de}}
\IEEEauthorblockA{\IEEEauthorrefmark{2}Fraunhofer IEE, Kassel, Germany. \emph{\{%
\href{mailto:philipp.haertel@iee.fraunhofer.de}{philipp.haertel},
\href{mailto:christoph.scholz@iee.fraunhofer.de}{christoph.scholz},
\href{mailto:malte.siefert@iee.fraunhofer.de}{malte.siefert},} \\
\emph{\href{mailto:sebastian.wende-von.berg@iee.fraunhofer.de}{sebastian.wende-von.berg}%
\}@iee.fraunhofer.de}}
\IEEEauthorblockA{\IEEEauthorrefmark{3}Technical University of Darmstadt, Germany. \emph{\href{mailto:florian.steinke@eins.tu-darmstadt.de}{florian.steinke@eins.tu-darmstadt.de}}
\IEEEauthorblockA{\IEEEauthorrefmark{1}Corresponding author}
}}

\maketitle

\begin{abstract}
	Future energy systems are subject to various uncertain influences.
	As resilient systems they should maintain a constantly high operational performance whatever happens.
	We explore different levels and time scales of decision making in energy systems, highlighting different uncertainty sources that are relevant in different domains.
	We discuss how the uncertainties can be represented and how one can react to them.
	The article closes by summarizing, which uncertainties are already well examined and which ones still need further scientific inquiry to obtain resilient energy systems.
\end{abstract}




\section{Introduction}
\label{sec:introduction}

Future sustainable energy systems based on renewable energy sources will have a significantly higher complexity due to a large number of actively integrated elements, weather dependencies, and dynamic interactions.
Increased uncertainties originating from weather conditions, energy user behaviors, investments and operational decisions of millions of actors, to name just a few sources, have to be addressed. 
Intelligent methods that enable flexible operations are circuital to handle complexity \cite{Loeser2022}.
Core elements are predictions under uncertainty (e.g., nowcasts or forecasts of influential parameters or a status) to enable decision making for operation and control of future energy systems or investments.
In the end, the quality of predictions and decisions will define performance and resilience of a sustainable energy system \cite{Resi2023}.    

\begin{figure}
	\includegraphics{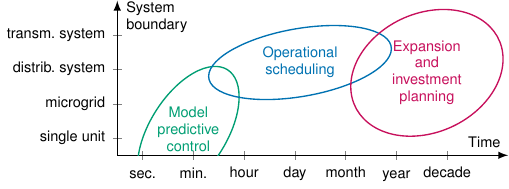}
	\caption{Time horizon and system boundaries of decision-making schemes.}
	\label{fig:plotTimeSize}
\end{figure}

Within this article, we will explore predictions and decision making at different system and time scales, highlighting relevant uncertainty sources.
We will discuss how they can be modeled and how resilient decisions can be derived in different areas (cf. \cref{fig:plotTimeSize}) including 
\ac{mpc} of small-scale systems (seconds to minutes),
operational scheduling of distribution and transmission systems (minutes to days), and
expansion and investment planning (years and decades).

The remainder of this article is structured as follows.
First, we will highlight recent trends in probabilistic predictions in \cref{sec:uncertaintyModeling}.
Then, we will discuss the use of different decision making schemes for different timescales and system boundaries, namely, \ac{mpc} in \cref{sec:mpc}, operational planning in \cref{sec:operationPlanning}, and expansion and investment planning in \cref{sec:expansionPlanning}.
\Cref{sec:conclusions} concludes the paper by summarizing which extreme cases are considered and which are still left out in today's decision making schemes.


\section{Recent trends in uncertainty modeling} 
\label{sec:uncertaintyModeling}

\begin{figure*}[ht!]
    \centering
    \includegraphics{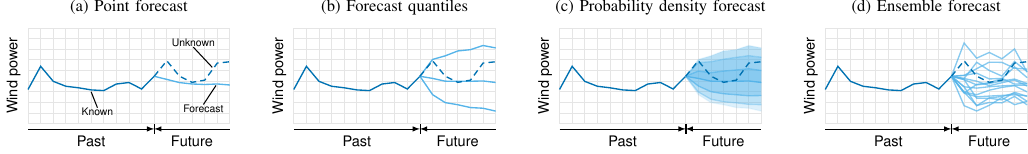}
    \caption{The dashed line is the (unknown) ground truth. In contrast to the point forecast, the probabilistic forecasts reflect the variability in future outcomes.
    Quantile and density forecasts directly describe the uncertainty in the forecast, while ensemble forecasts consider discrete future outcomes. Data from \cite{Han2021}.}
    \label{fig:forecasts}
\end{figure*}

In contrast to point predictions, which only predict a single outcome, probabilistic predictions quantify uncertainties in predicted outcomes.
Typical forms include quantiles, density predictions, or ensemble/scenario predictions \cite{haben2023,Gensler2017} (cf. \cref{fig:forecasts}).
They represent possible variants of an unknown present (nowcasts) or future (forecasts), which depend on non-deterministic influences, e.g., weather conditions or human behavior. 
Probabilistic forecasts are often updated in a sliding time window manner to obtain so-called rolling forecasts.
Moreover, several prediction steps within a given forecasting horizon are typically employed, e.g., using a one-hour steps size in a forecast for an entire day.
This allows decision processes to make better-informed choices.
%
This is particularly important, e.g., in reinforcement learning, where context resilience for decision making is crucial, because it allows the system to respond to different predictions and understand their impact on future outcomes.
Additionally, probabilistic forecasts enable decision schemes to anticipate unlikely events.


Various approaches to obtain probabilistic predictions exist.
Classical approaches include parametric models that assume a functional form of probability distributions, e.g., Gaussian or mixture models, quantile regression \cite{Bieshaar2020}, non-parametric techniques, e.g., kernel density estimation, or copula.
Recently, neural approaches have become popular.
Among them are long short-term memory \cite{Gensler2016}, Bayesian neural networks, generative adversarial networks \cite{janke2020probabilistic}, (variational) autoencoders \cite{Gensler2016}, and \acp{cnn}.


Probabilistic predictions are only suitable for reliable decision-making if they are well-calibrated, i.e., if the predicted uncertainty accurately approximates the true probability distribution of the predicted events. 
For quantile predictions, one must also ensure that the smaller quantiles do not exceed the bounds of the larger ones \cite{Decke2024}.

\begin{figure}[b]
    \centering
    \includegraphics{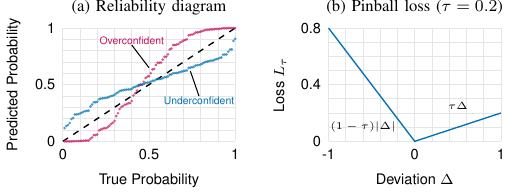}
    \caption{The reliability diagram plots observed quantiles against predicted ones. Deviations from the diagonal indicate over- or underconfidence.
    A reliable predictor should be close to the diagonal.
    The pinball loss penalizes the deviation $\Delta$ between a predicted quantile and an observed value.}
    \label{fig:error-measures}
\end{figure}

Different approaches to assess the reliability of prediction models exist, e.g., reliability diagrams (cf. \cref{fig:error-measures}a).
These allow to identify if a prediction is overconfident, i.e., it underestimates the actual uncertainty, or underconfident.
For further assessment, skill scores are often used.
These compare the relative accuracy of a predictor with a known benchmark~\cite{haben2023}.


To build reliable prediction systems it is necessary to calibrate them during training (or afterwards).
The pinball loss, for instance, quantifies the discrepancy between predicted quantiles and observed values (cf. \cref{fig:error-measures}b).
The continuous ranked probability score is another frequently used measure that allows to minimize the distance between the observed distribution and the predicted cumulative distribution function.
For a survey on different uncertainty assessment techniques, we refer to \cite{Gensler2018}.
There is also some research on use-case-dependent loss functions \cite{beykirch2024value}.

The availability of training data significantly influences performance (and potential model complexity).
Machine learning methods often require large amounts of training data to ensure reliable predictions by reducing the epistemic uncertainty of models. 
However, if there is a high aleatoric uncertainty in the data, i.e., uncertainty in the underlying process itself, for instance, load in the low-voltage grid \cite{haben2023}, then predictions will always have a high degree of uncertainty.

For many energy system applications, computational requirements are crucial.
Probabilistic predictions are often more complex than point predictions or pure statistical approaches, but there are also differences within these schemes:
A quantile regression network needs only one pass through the network to infer a forecast, while a variational autoencoder approach needs several passes.
The choice of the model thus also depends on time constraints such as real-time requirements.


\section{Model predictive control}
\label{sec:mpc}

As a control method relying on forecasts, \ac{mpc} \cite{GP2019} is widely employed in energy systems on different time-scales.
A possible objective is to provide setpoints to the storage units, conventional and renewable generators, and curtailable loads -- if present -- to operate systems in a safe and meaningful manner.
The general structure of \ac{mpc} schemes is as follows:
A forecasting scheme (see \cref{sec:uncertaintyModeling}) is used to obtain predictions of the uncertain inputs over a future time horizon.
For these forecasts and a measurement of the current system state, the \ac{mpc} solves an optimization problem online to determine the best future behavior with respect to a chosen objective.
The optimization iterates the control inputs over the future horizon and evaluates a dynamic model (subject to the uncertainties) to find an evolution that satisfies the control objectives.
The minimization of an objective function $\ell(\cdot)$ subject to constraints \eqref{eq:mpc:constraints} can be formulated as in \cref{prob:mpc}:
The contained variables can be partitioned
into control inputs $u(j)$,
system states $x(j)$,
algebraic auxiliary variables $z(j)$,
and uncertain inputs $w(j)$.
Moreover,
$f_{\mathrm{state}}$,
$f_{\mathrm{ineq.}}$, and
$f_{\mathrm{eqal.}}$ are functions used to implement constraints.
The discrete-time formulation uses $k$ to denote the current time and $j$ to denote a forecast step.

\begin{problem}[\acs{mpc}]
\label{prob:mpc}
	Solve the optimization problem
	\begin{align}
		\min_{u(k),\ldots,u(k+J-1)} ~ \textstyle\sum_{j = k+1}^{k+J} \ell\big(u(j-1), z(j), x(j)\big) \label{eq:mpc:objective}
	\end{align}
	subject to (for $j\in\{k,\ldots k+J-1\}$ and $x(k) = x_k$) :
	\begin{subequations}\label{eq:mpc:constraints}
		\begin{align}
			x(j+1) 	&= f_{\mathrm{state}} \big( x(j), u(j), w(j) \big), \label{eq:mpc:constraints:dynamics} \\
			0 		&\leq f_{\mathrm{ineq.}} \big( x(j+1), u(j), z(j+1), w(j) \big), \\
			0 		&= f_{\mathrm{eqal.}} \big(x(j+1), u(j), z(j+1), w(j)  \big).
		\end{align}
	\end{subequations}
\end{problem}

For a measured current state $x_k$, \eqref{eq:mpc:constraints} allows to model the future behavior of the system under control and subject to constraints such as, e.g., unit or line power limits.
This, of course, is only possible if the uncertain inputs $w(j)$ are forecast suitably.
If  $w(j)$ is forecast, the solution of \eqref{eq:mpc:objective} in $k$ results in an trajectory of inputs $u(j)$, of which typically the first value is applied to the power system, and the optimization is repeated in $k+1$  using updated forecasts and measurements, thus leading to a receding horizon control scheme.
This concept can be applied to various tasks in power systems, ranging on different time-scale from the sub-second domain to forecasts over several days (or even years in investment planning).
Likewise, $w(j)$ can encode fast transients, e.g., the effects of short-circuits or link failures {\cite{SHWS17,TS21}}, over weather forecasts for generation,  up to attack response for system resilience and restoration \cite{borner2021measuring}.
Two instances are detailed below.

\subsection{Dealing with external uncertainties}

In islanded power systems, uncertain renewable generation and loads have to be entirely compensated to ensure a local power balance.
Thus, effects such as smoothing of renewable generation and loads by geographical dispersion only come to bear in a very limited manner and large gradients in renewable generation and loads can appear.
Highly uncertain $w(k)$ which are expected in future energy systems are already present in these grids today. 
To deal with uncertain loads and renewable infeed, minimax \ac{mpc} that employs robust forecast intervals (see \cref{fig:forecasts}\,b) can be used.
In \cite{HNRR2014}, such an approach was compared to a scheme that employs a point forecast (see \cref{fig:forecasts}\,a).
It was shown that the minimax \ac{mpc} comes with increased robustness but also with quite conservative control actions.
To address this issues, stochastic \ac{mpc} schemes that consider discrete probability distributions in the form of ensembles (see \cref{fig:forecasts}\,d) and minimize the expected costs over all scenarios can be used \cite{HSB+2015}.
Such approaches tend to perform well, also in terms of robustness, if the forecast probability distributions are accurate.
However, incorrect forecast probabilities can lead to bad performance of aforementioned approach.
To counteract this effect, risk-averse approaches can be employed \cite{HSR+2020}.
These allow to consider uncertainty in forecast probabilities and to choose how much one trusts a prediction.
Effects associated with not-so-perfect forecasts can so be mitigated by choosing to trust less.
Moreover, these approaches allow to emphasize \ac{hilp} events, i.e., events that are very unlikely to happen but have a severe effect if they occur, which increases resilience.

\subsection{Dealing with uncertain system models}

In context of robust and resilient energy systems, the model \eqref{eq:mpc:constraints:dynamics} is often subject to simplifications or parametric uncertainties.
Reasons can be
that linearized variants of nonlinear processes are used,
that the model is learned from historic data (see, e.g., \cite{LH2024}), or
that the dynamics are affected by disturbances or faults.
For these cases, an extension of \eqref{eq:mpc:constraints:dynamics} to
\begin{align}
     x(j+1) 	&= f_{\mathrm{state}} \big( x(j), u(j), w(j),\theta(j) \big), \label{eq:mpc:constraints:dynamics:ext}
\end{align}
is constructive, where $\theta(j)$ subsumes all internal model uncertainties.
The requirement of robustly stabilizing control make it necessary to compute $u(j)$ from solving \eqref{eq:mpc:objective} such that the \ac{mpc} scheme converges for any possible realization of $\theta(j)\in\Theta_k$ from a bounded, possibly time-varying set $\Theta_k$.
Among different versions of robust \ac{mpc}, the use of \ac{lpv} models has proven useful for energy systems \cite{TSS16,SHWS17}.
Along this line, a combination of \ac{mpc} with embedded feedback controllers synthesized for \ac{lpv} models has been developed to control wind farms \cite{TS19}.
The extension to fault-tolerant control with respect to $\theta(j)$ representing larger faults in order to contribute to system-wide resilience is, however, still matter of future work.


\section{Operational scheduling}
\label{sec:operationPlanning}

The expansion of renewable energies and the associated transformation of the energy system are leading to increased uncertainties in operational scheduling, which affects the resilience of the energy system.
These uncertainties are mainly associated with loads and renewable generation and their forecasts \cite{saintdrenan2016uncertainty,saintdrenan2017probabilistic}, as well as uncertain measurements in the grids (or even nowcasts of grid states), and introduce risks for grid operation and flexibility scheduling.
Existing operational scheduling schemes do not fully account for these uncertainties.
Often, conservative safety margins are used to account for uncertainties and avoid issues such as equipment overloading, which results in inefficient usage of energy system capabilities.
Uncertainties also result in risks in grid automation, particularly in low voltage grids where hundreds of thousands of operational decisions are made each day.

Consequently, uncertainties and even \ac{hilp} events, e.g., asset faults, extreme weather conditions, or malicious attacks, must be explicitly considered in future operational scheduling schemes \cite{moretti2020efficient}.
This includes probabilistic state estimations, probabilistic forecasts of conditions, and optimization approaches that can handle uncertainties.
The outcome will likely be probability distributions, which must be integrated and processed in operational scheduling schemes.

Operational planning today still largely relies on point forecasts (cf. \cref{fig:forecasts}a) that predict expected values without accounting for uncertainties.
Although various probabilistic forecasting methods have been published and implemented \cite{bessa2017towards,dobschinski2017uncertainty}, many are not tailored to the specific needs of operational scheduling.
Probabilistic forecasts must be standardized and methods developed to derive meta-forecasts through fusion of information from multiple sources, to improve results.
There is also a lack of procedures for integrating complex probabilistic forecasts and associated decision-making schemes.
Additionally, understanding probabilistic information can be challenging for grid operators, especially under time pressure.

To enable probabilistic decision-making, uncertainties must be explicitly identified.
Associated risks should be assessed by evaluating potential consequences for grid operations.
Optimizations should then be carried out using appropriate metrics that enable resilient decisions.
Moreover, technical challenges such as handling large data volumes and complex calculations need to be addressed.
Furthermore, personnel must be trained in using probabilistic information, and appropriate decision support and visualization tools must be provided.

New methods are needed to consider uncertainties in system operations.
Ensemble forecasts (cf. \cref{fig:forecasts}d) and measurement error distributions lay the foundation for future approaches.
In \cite{brendlinger2024} an ensemble forecast is used for grid state estimation by producing probability distributions of, e.g., line loading.
In \cite{wang2022robust} it is demonstrated how to consider uncertainties in flexibility estimations for high voltage grids.
With such an approach, optimization results can be obtained that are resilient against unexpected changes.
In order to use aforementioned approaches in real-time decision-making, fast artificial intelligence-based optimization can be employed.

If uncertainties are taken into account in decision-making, then systems can become more resilient and can be operated closer to and safer within capacity limits.
In order to do so, it is important to reduce forecast errors and also include possible \ac{hilp} events in estimations for $N-k$ contingency analysis.
However, uncertainties will always remain, and grid operators must learn to cope with them.
There are methods and approaches to harness these uncertainties and risks, which are an important prerequisite for operating the grid in a more resilient manner.
In addition, there is a need for an adaptation of market mechanisms that enable short-term reactions and make use of market flexibility to support grid stability \cite{He2020}.
The development of probabilistic decision strategies and risk management will lead to increased resilience in case of faults, disturbances, and malicious attacks.


\begin{table*}[t]
\caption{Consideration of uncertainties in different domains.}
\label{tab:summary}
\begin{tabularx}{\textwidth}{>{\raggedleft\arraybackslash}p{14.4mm}>{\raggedright\arraybackslash}X>{\raggedright\arraybackslash}X>{\raggedright\arraybackslash}p{75mm}}
	\toprule
	& Model predictive control & Operation planning & Expansion and investment planning \\ 
	\midrule
	Often considered uncertainties / extreme cases & 
		\begin{compactitem}
			\item Weather conditions \newline (incl. \textit{dunkelflaute})
			\item Demand conditions
			\item Price in energy markets
		\end{compactitem}
            & \begin{compactitem}
			\item Weather variability \newline (incl. \textit{dunkelflaute})
            \item Demand conditions
            \item Price in spot markets
            \item Outages (planned and unplanned)
		\end{compactitem}
            & \begin{compactitem}
			\item Weather variability (incl. \textit{dunkelflaute})
            \item Outages (planned and unplanned)
            \item Commodity prices
            \item Technological developments
            \item Policies, measures, instruments, arrangements
            \item Existing infrastructure planning
		\end{compactitem}\\
	\midrule
	Rarely considered uncertainties / extreme cases & 
		\begin{compactitem}
			\item Loss of power lines
			\item Loss of entire units
			\item Short circuits
			\item Large overshoots at the lower control layers
		\end{compactitem}
	& 
		\begin{compactitem}
			\item Intentional physical damage
            \item Malfunction of information and communication technology
			\item Malicious attacks
		\end{compactitem}
            & \begin{compactitem}
			\item Climate change hazards, especially compound events
            \item Geo-political events, e.g., wars, terrorism, sabotage
            \item Availabilities of raw materials, strategic technologies, goods 
            \item Social acceptance conflicts, individual behavior, such as generation and load placement
		\end{compactitem} \\
	\bottomrule
\end{tabularx}
\end{table*}

\section{Expansion and investment planning}
\label{sec:expansionPlanning}

Expansion and investment planning range from constructing transformation pathways for complete multi-energy systems of countries over a horizon of several decades \cite{howells2011osemosys} to specific technological investment decisions, e.g., about required measurement and control sites in a given distribution grid \cite{mora2021minimal}, the placement of individual units, e.g., \acl{facts} devices, in transmission grids \cite{santos2023robust}, or the replacement of units close to the end of their expected lifetime \cite{huang2024spatial}.
In all cases, models must endogenously capture interactions and trade-offs between investment decisions and their operational use, together with their overall system consequences.

Variability in day-to-day operational use is typically considered in the form of representative time periods, e.g., relying on historic load profiles or time series of available renewable generation \cite{Henze2020}.
The operational reaction of the system design typically comes from deterministic optimization \cite{howells2011osemosys,Frischmuth.2024b}, meta-heuristic approaches \cite{Scheidler.2018}, optimization under uncertainty with various forms of risk operators \cite{ROALD2023108725}, or it is summarized into closed-form control policies \cite{mora2021minimal,santos2023robust}.
The typical design objective is to minimize the expected total costs, considering both investment and operational factors.
However, one can also consider worst-case costs or regret, or the so-called (conditional) value at risk \cite{ackermann2022design}, which mixes the two extremes and leads to risk-averse optimization.

Planning resilient energy systems goes beyond capturing load and generation variability, involving complex system interactions and uncertainties.
Integrated systems may increase redundancy with diverse technologies and limit spatial risks by decentralization.
However, they also increase dependence on electricity infrastructure and bear the risk of disturbances spreading across sectors.
Transformation pathways are influenced by policy decisions, societal events, e.g., wars or technology developments, and other disruptive developments.
A common way to deal with these uncertainties is to consider a small set of hand-crafted scenarios and optimize the system independently for each one.
However, more systematic, model-endogenous approaches allow to better address resilience aspects, e.g., to handle the impacts of climate change through short- and long-term meteorological variability \cite{Schmitz.2024} or combine weather variations with uncertain industry demands~\cite{Frischmuth.2024a}.
Additionally, future spatial allocation of renewable units and flexibility installations depends on user acceptance which is difficult to model, and existing infrastructure details, especially in low-voltage grids, are often imprecise.
Here, probabilistic reconstruction approaches may offer a solution \cite{tomstumetste24}.

Recourse decisions, e.g., shedding loads~\cite{Frischmuth.2024a} or overload protection schemes \cite{Gourab.2023}, and their associated costs are pivotal in system expansion planning, ensuring that the systems can respond effectively to the myriad of uncertainties that they face.
These differ widely: some have distributions that are known or can be estimated from data, others have only bounds on the range of possible values, and some are characterized by unforeseeable events and emergent scenarios.
The latter poses a significant challenge as it necessitates designing adaptive systems that can withstand any kind of unforeseen variability at run-time.
Currently, however, it remains unclear how such resilience can be achieved.


\section{Conclusions}
\label{sec:conclusions}

In this article, we discussed how resilience can be ensured in different areas of future energy systems.
It becomes obvious that resilience must be taken into account already in the design process of intelligent systems right from the development phase \cite{Tomforde2019}.
\Cref{tab:summary} presents an overview of considered and left-out extreme cases.
Based on this overview, multiple future research directions can be identified which guide the way to resilient intelligent sustainable energy systems.

\bibliographystyle{IEEEtran}
\bibliography{literature}

\end{document}